# Two-dimensional finite element simulation of fracture and fatigue behaviours of alumina microstructures for hip prosthesis

Kyungmok Kim[1], Bernard Forest[1], Jean Geringer[1*]

[1]Ecole Nationale Supérieure des Mines de Saint-Etienne, ENSM-SE,

Center for Health Engineering, UMR CNRS 5146, IFR 143

Department of Biomechanics and Biomaterials,

158 cours Fauriel, F-42023 Saint-Etienne Cedex 02, France

Tel: +(33).4.77.42.66.88; geringer@emse.fr

**Abstract**

This paper describes a two-dimensional finite element simulation for fracture and fatigue behaviours of pure alumina microstructures found at hip prosthesis. Finite element models are developed using actual $Al_2O_3$ microstructures and a bilinear cohesive zone law. Simulation conditions are similar to those found at a slip zone between a femoral head and an acetabular cup of hip prosthesis. Effects of microstructures and contact stresses are investigated in terms of crack formation. In addition, fatigue behaviour of the microstructure is determined by performing simulations under cyclic loading conditions. It is identified that total crack length observed in a microstructure increases with increasing the magnitude of applied contact stress. Cyclic simulation results show that progressive crack growth occurs with respect to number of fatigue cycles. Finally, this proposed finite element simulation offers an effective method for identifying fracture and fatigue behaviours of a microstructure.

---

[*] Corresponding author. Email address: geringer@emse.fr

## 1. Introduction

Alumina is known as a primary ceramic material in biomedical industry. Particularly, $Al_2O_3$ is widely used as the material of a femoral head and an acetabular cup of hip prosthesis. $Al_2O_3$ maintains good biocompatibility, high mechanical strength and fracture toughness, showing low friction coefficient at the contact surface of hip prosthesis. Alumina ceramics are available in different purities, typically from 85 % to 99.5 %, and high purity (> 99.5 %) alumina is used in hip prostheses [1, 2]. Repeated mechanical loadings, due to human gait and micro separation between head and cup, are imposed on the surface between a head and a cup of artificial hip prosthesis, leading to shock degradations [3]. Failure of hip prosthesis resulting from shocks brings about serious damages to the human body. For this reason, shock degradation is one of critical concerns in design of hip prosthesis. Several studies were carried out in order to investigate mechanical and wear damages of hip prostheses [4-6]. Hausselle [4] investigated fracture toughness and wear rate of alumina heads or cups of hip prostheses by performing shock experiments. Stewart et al. [5] investigated wear and fracture of alumina cups against zirconia heads with a hip simulator. Wear rate was measured under micro-separation conditions. De Aza et al. [6] determined crack growth resistance of bio-ceramics including alumina applicable to hip prostheses.

Finite element method is widely used for simulating crack initiation and propagation of a structure resulting from external loading. Warner and Molinari [7] developed a two-dimensional finite element model of compressive fracture in ceramics. Finite element modelling was performed with pure alumina microstructures generated by Voronoi tessellation of randomly positioned seeds. Sfantos and Aliabadi [8] performed micro-and macro-scale boundary element modelling with polycrystalline $Al_2O_3$ ceramics. Three-point bending was loaded to a model and crack propagation was then investigated. However, a finite element model using actual alumina microstructures for hip prosthesis has not been developed in spite of its importance.

Several methods for investigating failure behaviour have been considered including fracture mechanics, continuum damage mechanics, and so on. Particularly, a method using cohesive zone law is remarkable, since this method enables simulating fracture of interfaces between physical parts and characterizing post-yield softening. Elements used in cohesive zone modelling do not represent any physical material but contain cohesive forces arising when they are being pulled apart. If these elements satisfy a pre-defined damage criterion, cohesive forces are completely removed. Thus, it is possible to simulate crack growth in a structure. Camacho and Ortiz [9] developed a lagrangian finite element method of fracture and fragmentation in brittle materials. Propagation of multiple cracks under impact loading was modelled with a cohesive zone law. Thermal effect induced in the course of impact loading was taken into account in the model. Ortiz and Pandolfi [10] extended this cohesive model to three dimensions. A three-dimensional finite-deformation cohesive element was developed, using irreversible cohesive laws. Nguyen et al. [11] developed a cohesive model for fracture and fatigue behaviours of a plane strain sample. An irreversible cohesive law with unloading-reloading hysteresis was implemented for describing fracture processes. Espinosa and Zavattieri [12, 13] developed a grain level model for failure initiation and evolution in polycrystalline brittle materials. Wave propagation experiments were simulated with various properties of cohesive zone [12]. In addition, finite element analysis of ceramic microstructures subjected to dynamic pressure-shear loadings was performed with cohesive zone law [13]. Yang et al. [14] formulated a cohesive fracture model for human femoral cortical bones. Subit et al. [15] developed a micro-mechanical model for predicting damages in ligament-to-bone attachment of a human knee joint. A cohesive zone model theory was proposed, focusing on the development of behaviour laws for crack initiation and propagation at an interface within a fibrous material or at the interface between materials.

In fully dense $Al_2O_3$ ceramics, grains are rather various in size and shape. Thus, difficulties in modelling $Al_2O_3$ microstructures exist, leading to the increase of simulation time. A commercial finite element package (ABAQUS® 6.8) enables building complex elements and implementing cohesive zone models. In this paper, a two-dimensional finite element model was developed by using actual $Al_2O_3$ microstructures. Fracture behaviour of grain boundaries was described with a bilinear cohesive law. Simulation conditions chosen were similar to those found at hip prosthesis. Crack propagation in $Al_2O_3$ microstructures was investigated, and fatigue behaviour of the microstructure was then obtained by performing simulations under cyclic loading conditions.

## 2. Cohesive model

Mechanical properties of grain boundaries can be described with a bilinear, time-independent cohesive zone law. Cohesive behaviour is directly defined in terms of a traction-separation law [16]. This cohesive behaviour allows specification of mechanical properties such as relative displacement at failure, stiffness, and strength. Moreover, the behaviour allows assumption that the failure of elements is characterized by progressive degradation of the material stiffness.

A cohesive element is subjected to normal and shear displacements under loading condition. The maximum value of a displacement is defined by $\delta^{max} = \sqrt{\langle \delta_n^{max} \rangle^2 + (\delta_s^{max})^2}$, where $\delta_n^{max}$ and $\delta_s^{max}$ are maximum values of normal and shear displacements attained during the loading history, respectively. In order to quantify the damage of a cohesive element, a damage variable ($D$) proposed by Camanho and Davila [17] is used

$$D = \frac{\delta^f (\delta^{max} - \delta^0)}{\delta^{max}(\delta^f - \delta^0)} \tag{1}$$

where $\delta^f$ denotes the effective displacement at complete failure, and $\delta^0$ is the effective displacement when normal stress ($T_n$) and shear stress ($T_s$) of a cohesive element satisfy the following equation.

$$\left(\frac{\langle T_n \rangle}{T_n^{max}}\right)^2 + \left(\frac{T_s}{T_s^{max}}\right)^2 = 1 \qquad (2)$$

where $T_n^{max}$ and $T_s^{max}$ are maximum values of the nominal normal stress and the nominal shear stress, respectively. The symbol $\langle \rangle$ denotes that a pure tensile deformation initiates damage.

[Fig. 1]

Fig. 1 illustrates the stress versus displacement curve for cohesive behaviour. Normal and shear displacements are considered in cohesive elements. Cohesive behaviour under pure tensile deformation is described in Fig. 1b; line 1 is a loading and unloading path before damage initiation. Line 2 is an example of an unloading and reloading path after damage initiation. Cohesive behaviour in the shear direction is similar to that in the normal direction. Normal stress ($T_n$) and shear stress ($T_s$) can be expressed as

$$T_n = \begin{cases} (1-D) \times \overline{T}_n, & \overline{T}_n \geq 0 \\ \overline{T}_n, & \overline{T}_n < 0 \end{cases}$$
$$T_s = (1-D) \times \overline{T}_s \qquad (3)$$

where $\overline{T}_n$ and $\overline{T}_s$ are normal and shear stresses calculated by the elastic traction-separation behaviour for the current strains without damage, respectively. The cohesive model described above can be implemented in a finite element framework. From the principal of virtual work, an equilibrium form is expressed as

$$\underbrace{\int_\Omega \mathbf{S}:\delta\varepsilon d\Omega}_{internal} - \underbrace{\int_{\Gamma_T} \mathbf{T}_{ex}\cdot\delta\mathbf{u}d\Gamma_T}_{external} - \underbrace{\int_{\Gamma_C} \mathbf{T}\cdot\delta\Delta d\Gamma_C}_{cohesive} = 0 \qquad (4)$$

where $\mathbf{S}$ and $\varepsilon$ are internal stress and strain tensors. $\Omega$, $\Gamma_T$ and $\Gamma_C$ denote the volume, external boundary, and cohesive boundary, respectively. $\mathbf{T}_{ex}$ is the externally applied traction, and $\mathbf{u}$ and $\Delta$ are displacement vectors. $\mathbf{T}$ is a cohesive traction tensor. The last term in equation 4 is equal to the virtual work done by cohesive elements.

For the purpose of resolving the cohesive zone accurately, sufficient number of cohesive elements is needed. An approximate cohesive zone length ($L_{cohesive}$) under plain strain condition is given by [18]

$$L_{cohesive} = \frac{\pi \cdot E}{2(1-\nu^2)} \frac{G_{IC}}{(T_n^{max})^2} \qquad (5)$$

where $E$ is elastic modulus, $\nu$ is Poisson's ratio, and $G_{IC}$ is the cohesive energy. For ensuring mesh independency, the cohesive element size in the authors' simulation needs to be less than $\frac{L_{cohesive}}{6}$.

## 3. Finite element simulation

[Fig. 2]

Fig. 2 illustrates a proposed simulation algorithm for calculating crack lengths in a microstructure. A data digitalization program written by Python programming language digitalizes actual 2D microstructure images and generates finite element models including cohesive layers. Material properties of grains and cohesive layers are then defined. Plan strain, implicit analysis is performed in the simulation. During simulation, stresses and strains of all elements are calculated along with damage variables of cohesive elements. Fracture of grain boundaries is evaluated in terms of damage variables of cohesive elements. When the damage variable of an element is equal to unity, the element is deleted and considered as crack. Commercial finite element software (ABAQUS® 6.8 standard) was used for simulating models. Each model was simulated on a cluster with 10 calculation nodes (Intel Xeon Quad Core 3 GHz, 64 bits with 1 GB). Viscous stabilization was implemented to avoid severe convergence arising from softening behaviour and stiffness degradation, as recommended by ABAQUS [16, 17]. Dissipated energy fraction specified for stabilization was 0.002~0.005 during calculation.

[Fig. 3]

Fig. 3a and 3c show the microstructure images of $Al_2O_3$ (AKP-53, Sumitomo Co., Japan) sintered at 1400 °C for 2 hours. Fig. 3b illustrates a generated model with Fig 3a, including 23 grains. Dark lines in the model denote cohesive layers. The size of a quadrangle element for a cohesive element is 0.01 μm × 0.01 μm. All grains were meshed sufficiently finely for guaranteeing appropriate degree of resolution of local stress concentration effects. Fig. 3d is the other model generated by the same procedure.

These models are assumed to represent small volume elements at the contact surface of a head. During human gait, a head comes in frictional contact with the inner surface of a cup and then slips over the surface. It can be understood that some parts of microstructures located at the surface of a head are subjected to compressive and shear deformation resulting from sliding. In order to reproduce this deformation approximately, normal and shear stresses were applied to the top surface of a model. The bottom surface was fixed in all directions but the left and right sides remained free while the stresses were being applied. In these models, surface roughness was ignored, since it was less than 0.003 μm. The magnitude of normal stress varies according to locations within an actual contact surface, ranging from 5 MPa to 500 MPa [4]. In this paper, four normal stress magnitudes of 70, 80, 90 and 100 MPa were selected. According to the normal stress, the magnitude of shear stress was determined on the basis of Coulomb friction coefficient of the material (i.e. 0.4 at dry conditions and at room temperature) [19, 20].

[Table 1]

Table 1 shows mechanical properties of grains and cohesive layers used for this simulation. The maximum normal stress chosen lies within the range of 0.95-1.5 GPa found at literature [8, 21-23]. Under this condition, fracture energy release rate is equal to 1 $Jm^{-2}$ for a pure tensile mode and similar to those found at literature [24]. This fracture energy release rate was uniformly distributed to all cohesive layers. The maximum shear stress is assumed on the basis of the grain boundary shear strength's dependence on the compressive yield strength (3000 MPa) similar to friction coefficient (0.13) of a metallic glass [25]. Stiffness values of the cohesive element are assumed as $3.74 \times 10^{10}$ $MPa.mm^{-1}$ (in the normal direction) and $1.53 \times 10^{10}$ $MPa.mm^{-1}$ (in the shear direction).

## 4. Results and discussion

Contact stress between a head and a cup of hip prosthesis is one of key parameters inducing cracks in $Al_2O_3$ microstructures. For the purpose of investigating the effect of contact stresses, finite element simulations were performed with four different stress magnitudes.

[Fig. 4]

Fig. 4 shows the total crack length versus applied normal stress chart for models A and B after one loading-unloading cycle. Total crack length on the chart is defined as the sum of longitudinal distances of cracks generated in the microstructure. It is apparent that the total crack length rapidly increases with increasing applied stress magnitude. The crack growth rate of model A is higher than that of model B. In addition, the total crack occurred in model A is longer than that found in model B.

[Fig. 5]

Fig. 5 illustrates crack distribution of model A with respect to applied contact stress magnitude. 2D plots were obtained after one loading-unloading cycle. At a normal stress of 70 MPa, three cracks occurred within the entire model (Fig. 5a). At 80 MPa, longer cracks were observed at the upper-right side of a model (Fig. 5b). It was identified from the plots that the total crack length was increased with increasing applied contact stress magnitude (Fig. 5c). At 100 MPa, four grains located at the upper-right side of the model were finally separated from the microstructure, since cohesive layers surrounding the grains were completely removed (Fig. 5d). In addition, other cohesive layers were damaged due to the contact stress.

[Fig. 6]

Fig. 6 shows crack distributions of model B with respect to applied contact stress magnitude. These crack distributions were obtained after a loading-unloading cycle. Cracks initiated at the upper-left side of the model, differently from the result of model A. This crack path may be closed due to the applied normal stress but it is opened after unloading. Cracked zones were extended with increasing contact stress magnitude. At a normal stress of 100 MPa, long cracks occurred at the upper side of the model, eventually leading to grain separation from the microstructure.

[Fig. 7]

Cyclic loadings were applied onto model A for investigating fatigue behaviour of a microstructure. The multi-step finite element simulations were performed by applying contact stress in a triangular waveform. Fig. 7 illustrates crack distribution of model A with respect to number of fatigue cycles. Three small cracks occurred at the upper-right side of the model after the initial fatigue cycle. These cracks grew with increasing number of fatigue cycles. Long cracks were then generated by combining small ones. Additional cracks were also found at other grain boundaries, and these cracks gradually grew along deformed boundaries. Finally, cohesive elements surrounding three grains were almost removed. This progressive damage was allowed, since a stress field in a model was changed according to number of cycles. That is, stress concentration occurred near cracked areas after the initial cycle and cracked areas were exposed to larger displacement upon reloading. Furthermore, stiffness of a cohesive element was degraded when the stress reached a maximum value ($T^{max}$). The degradation of stiffness also increased the progressive damage. Crack path shown in Fig. 7c (after 15 cycles at a normal stress of 70 MPa) is similar to that found at Fig. 5c (after one cycle at a normal stress of 90 MPa). It can be concluded that repeated contact stresses lead to progressive damages in a microstructure, eventually resulting to grain separation.

Material loss rate in a microstructure can be approximately evaluated with the amount of separated grains with respect to number of cycles. If the amount of cohesive elements remaining between grains is so small that grains are considered to be separated in Fig. 7d, the loss area in model A is approximately $2.0\times10^{-6}$ mm$^2$ after 25 cycles. Supposing that grain thickness is 0.001 mm, the loss rate is calculated as $0.08\times10^{-9}$ mm$^3$/cycle. This loss rate corresponds to a local wear rate at contact surfaces of hip prosthesis subjected to high contact stress, since an applied normal stress of 70 MPa is approximately 12 times higher than a global value (5 MPa) found to contact between a head and a cup [4]. Thus, the total wear volume (~$8.0\times10^{-3}$ mm$^3$ per cycle) predicted in an entire contact area (~300 mm$^2$) is somewhat larger than the value (~$1.3\times10^{-3}$ mm$^3$ per cycle, i.e. 30 mm$^3$ in a head after 22500 cycles) measured with non-hot isostatically pressed alumina of hip prosthesis at dry condition [26].

**Commentaire [g1] :** Did you check the remark from Pr. Forest?
I agree with your number.

[Fig. 8]

[Fig. 9]

Fig. 8 shows the evolution of the total crack length with respect to number of fatigue cycles in model A. The total crack length gradually increases with increasing number of fatigue cycles and can be expressed as a linear function of a cycle. The slopes on the graph are equal to crack growth rates. It is identified from the slope values that crack growth rates of the model are similar within the range of 70-90 MPa despite small variance. Fig. 9 shows a graph of the magnitude of a cyclic normal stress (S) against the number of cycles to failure ($N_f$) of model A. The number of cycles to failure was defined as the cycle when the total crack length reached 6.3 μm. The number of cycles to failure was decreased with increasing the magnitude of the cyclic normal stress. Although further simulations under various loading conditions and microstructures are needed, this proposed simulation enables producing the S-N curve of a microstructure. The S-N curve can be used as a reference for predicting the lifetime of a microstructure.

In this simulation, external stresses were applied to the models for reproducing deformation of a microstructure at the contact surface of a head. Displacement could be imposed to the models instead of the shear stress. This modelling can be achieved after identifying actual relative displacement between a head and a cup. The number of grains in the selected models was 23 and 25, respectively. Thus, the selected models may not correspond to representative volume elements (RVE). Bigger models need to be generated and analyzed in future work. Nevertheless, it was demonstrated that this proposed method allows simulating fracture and fatigue behaviours of a microstructure.

## 5. Conclusions

This paper developed a two-dimensional finite element modelling for investigating fracture and fatigue behaviours of $Al_2O_3$ ceramics at the microscopic level (i.e. ceramic grain scale). Two different $Al_2O_3$ microstructures without voids were selected and modelled with commercial finite element software (ABAQUS®) and a data digitalization program developed by authors. Simulation algorithm from model generation to analysis was detailed, enabling cyclic loadings. A bilinear, time-independent cohesive zone law was implemented for describing fracture behaviour of grain boundaries. The cohesive zone law allows elements in grain boundaries to be removed when they satisfy a pre-defined failure criterion. Repeated mechanical loading gives rise to degradation of the contact surface between a femoral head and an acetabular cup of hip prosthesis. At the microscopic level, short and long cracks are observed among grain boundaries at the contact surface. The fracture phenomenon of a microstructure was reproduced by applying contact stresses to a model.

The effect of contact stresses was investigated, by applying four different stress magnitudes. Total crack length occurring in a microstructure was measured after an initial fatigue cycle (loading and unloading). It was identified that the crack length was increased with increasing the contact stress. In addition, differences of the crack length and path were observed between selected microstructures. It can be concluded that grain arrangement and size determine the total crack length and crack path.

It is of importance that fatigue behaviour of $Al_2O_3$ ceramic is investigated at the microscopic level. For the purpose of investigating the fatigue behaviour, cyclic loading was applied to a model. Short cracks were found at a variety of grain boundaries, and crack growth was apparently observed. That is, cracks progressively grew with increasing number of fatigue cycles. Results also showed that some grains were separated after the final fatigue cycle.

In conclusion, the proposed finite element model allows simulating fracture and fatigue behaviours of alumina microstructures found at hip prostheses. Further work will focus on the generation of a representative volume element, the development of a three-dimensional finite element model, and application to microstructures of bio-materials such as zirconia and alumina-zirconia ceramics.

**Acknowledgements**

The authors wish to acknowledge the financial support of 'ANR', related to the 'Opt-Hip' project. Particular thanks are due to Prof. Jerome Chevalier (INSA, France) for fruitful discussions.

alumina-alumina hip joints under standard and severe simulator testing conditions Biomaterials 2001, 22, 2191-2197.

**Figure Captions**

**Fig. 1** Illustration of cohesive response (a) in a mixed mode and (b) in a pure tensile mode.

**Fig. 2** Simulation algorithm. $N_i$ denotes the $i^{th}$ cycle.

**Fig. 3** Images of $Al_2O_3$ microstructures (purity of 99.99 %, and sintered at 1400 °C for 2 hours) and finite element models. The right images (b, d) are models generated with the left images (a, c). (b) and (d) are denoted as model A and model B, respectively.

**Fig. 4** Total crack length versus applied normal stress chart of models A and B. Shear stresses were applied with the normal stresses together.

**Fig. 5** Illustrations of crack distributions in model A with respect to applied contact stress magnitude: (a) a normal (shear) stress of 70 MPa (28 MPa), (b) a normal (shear) stress of 80 MPa (32 MPa), (c) a normal (shear) stress of 90 MPa (36 MPa), and (d) a normal (shear) stress of 100 MPa (40 MPa). Grey areas are $Al_2O_3$, black lines denote cohesive layers, and white ones denote cracks.

**Fig. 6** Illustrations of crack distributions in model B with respect to applied contact stress magnitude: (a) a normal (shear) stress of 70 MPa (28 MPa), (b) a normal (shear) stress of 80 MPa (32 MPa), (c) a normal (shear) stress of 90 MPa (36 MPa), and (d) a normal (shear) stress of 100 MPa (40 MPa). Grey areas are $Al_2O_3$, black lines denote cohesive layers, and white ones denote cracks.

**Fig. 7** Illustrations of crack distributions in model A with respect to number of fatigue cycles: (a) cycle 1, (b) cycle 10, (c) cycle 15, and (d) cycle 25. Grey areas are $Al_2O_3$, black lines denote cohesive layers, and white ones denote cracks. Triangular load wave was applied, ranging from zero to 70 MPa for normal stress (28 MPa for shear stress).

**Fig. 8** Evolution of the total crack length with respect to number of fatigue cycles in model A.

**Fig. 9** S-$N_f$ curve for model A. Circle markers are calculated data. The number of cycles to failure was defined as the cycle when the total crack length reached 6.3 μm.

**Table**

| Grain | | Cohesive layer | |
|---|---|---|---|
| Elastic modulus, E (shear, G), MPa | Poisson ratio, ν | Nominal normal stress, $T_n^{max}$ (Nominal shear stress, $T_s^{max}$), MPa | Displacement at failure, $\delta^f$, μm |
| 374000 (153000) | 0.22 | 1000 (400) | 0.002 |

Table 1. Mechanical properties of grains and cohesive layers.

**List of notation**

$a$: total crack length

$D$: damage variable

$E$: elastic modulus

$G$: shear modulus

$G_{IC}$ : cohesive energy

$L_{cohesive}$ : approximate cohesive zone length

$N$: number of cycles

$P$: applied normal stress

$T_n$: normal stress

$T_s$ : shear stress

$T_n^{max}$ : nominal normal stress

$T_s^{max}$ : nominal shear stress

$\bar{T}_n$ : normal stress calculated by the elastic traction-separation behaviour for the current strains without damage

$\bar{T}_s$ : shear stress calculated by the elastic traction-separation behaviour for the current strains without damage

**S**: internal stress tensor

**T**: cohesive traction tensor

**T**$_{ex}$ : the externally applied traction

**u**: displacement vector for external loading

$\delta^{max}$ : the maximum value of a displacement

$\delta_n^{max}$ : the maximum value of normal displacement during the loading history

$\delta_s^{max}$ : the maximum value of shear displacement during the loading history

$\delta^f$ : the effective displacement at complete failure

$\delta_n^f$ : the effective normal displacement at complete failure

$\delta^0$ : the effective displacement at damage initiation

$\varepsilon$ : strain tensor

$\Omega$ : volume

$\Gamma_T$ : external boundary

$\Gamma_C$ : cohesive boundary

$\Delta$ : displacement vector for cohesive element

$\nu$ : Poisson's ratio

Figure 1:

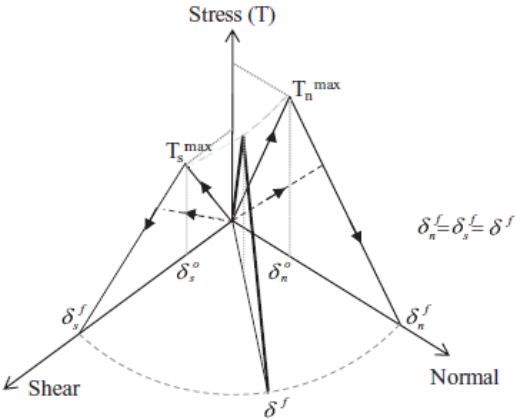

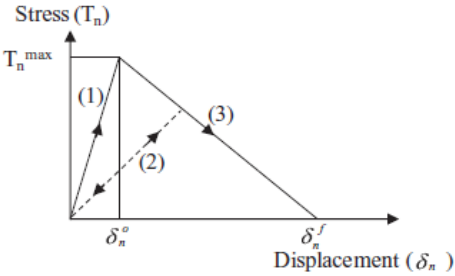

Figure 2:

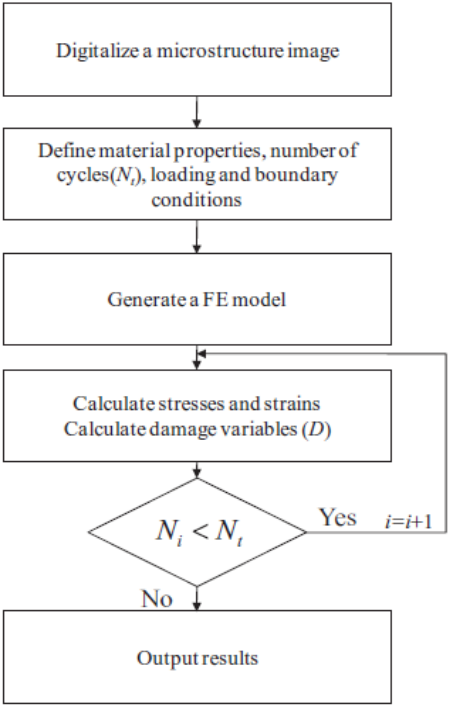

Figure 3:

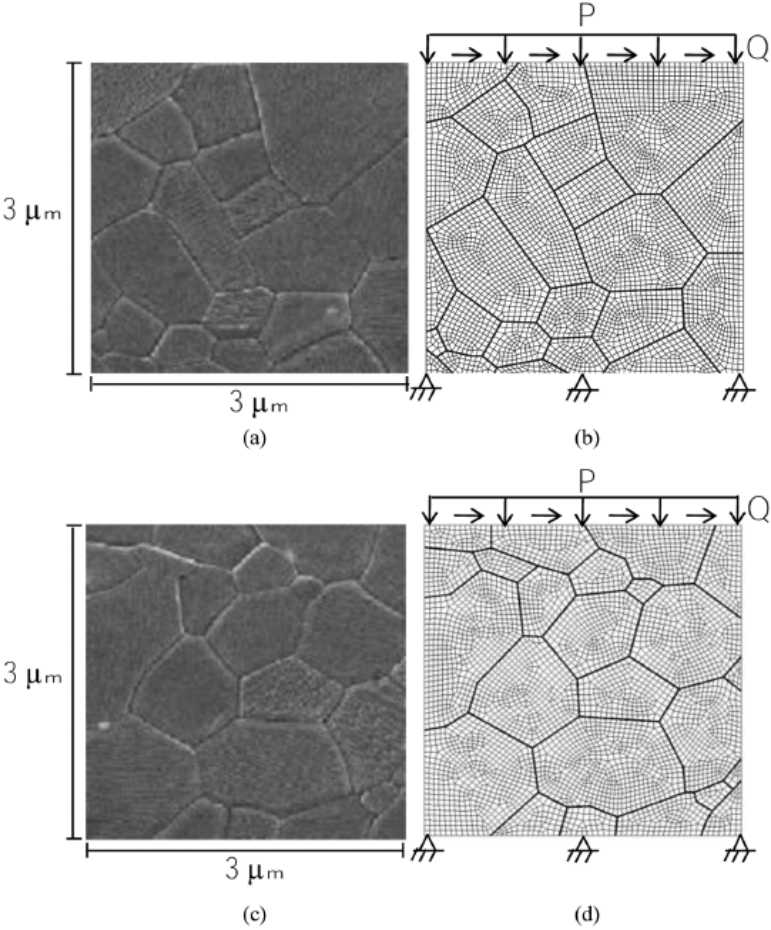

Figure 4:

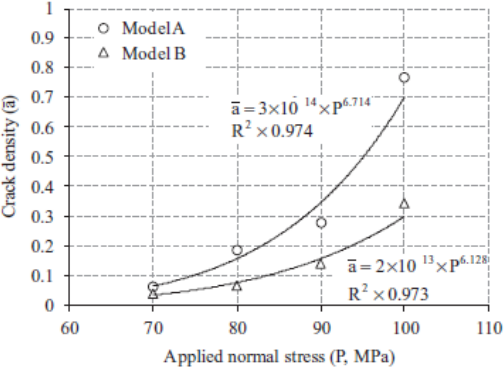

Figure 5:

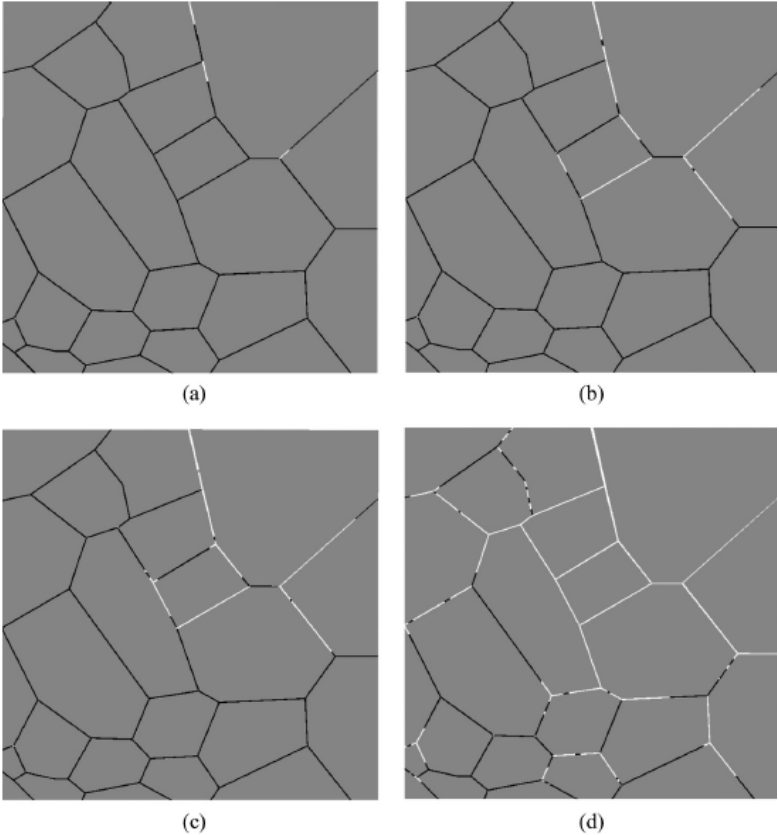

Figure 6:

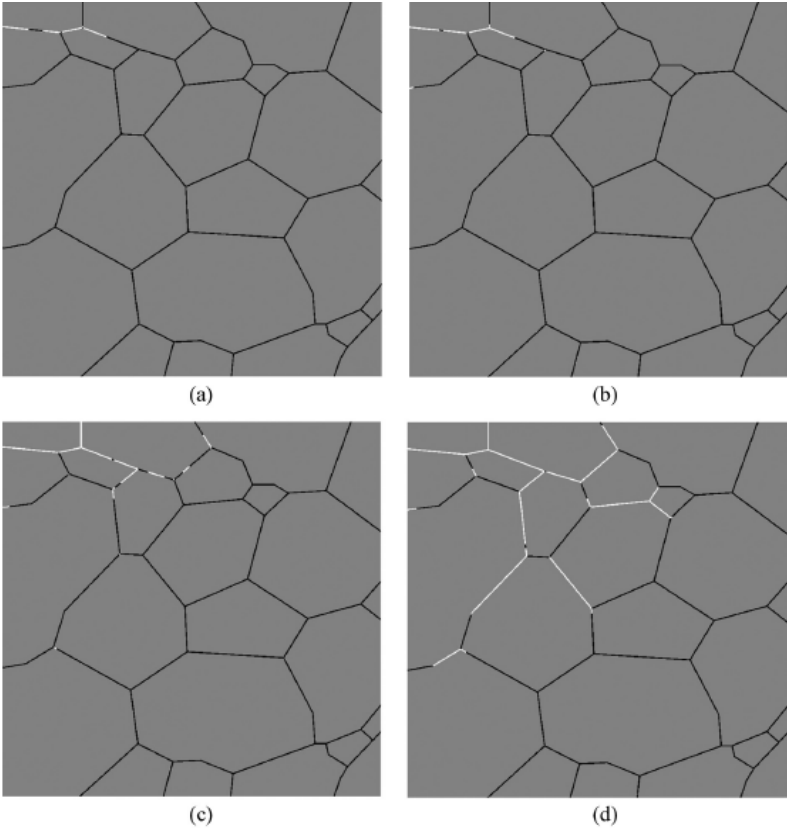

Figure 7:

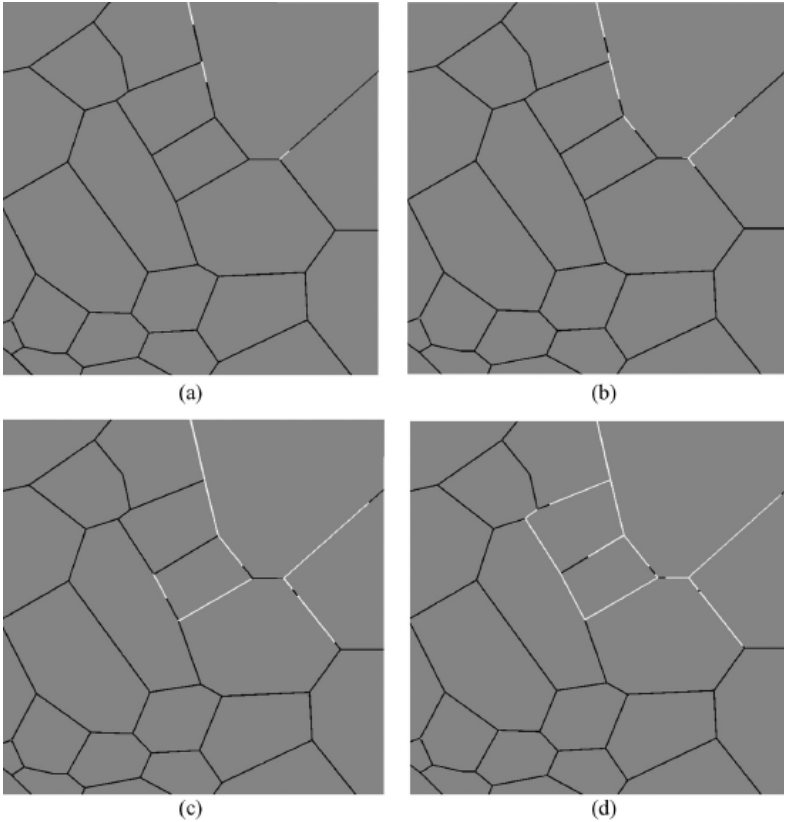

Figure 8:

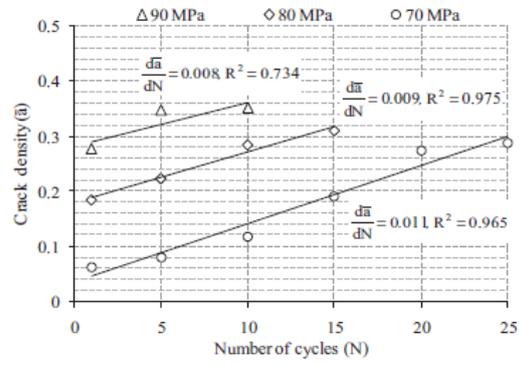

Figure 9:

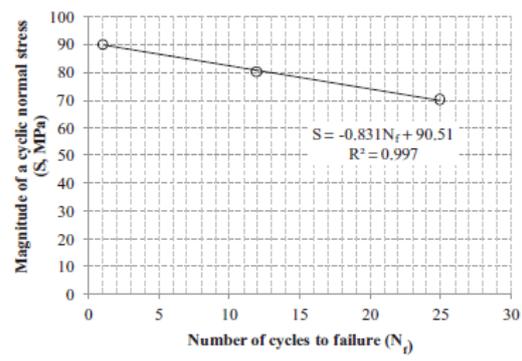